\documentstyle[amssymb,prl,aps,epsfig]{revtex}

\begin{document}

\title{Comment on "Correlation Between Superfluid Density and $T_{c}$ of Underdoped
YBa$_{2}$Cu$_{3}$O$_{6+x}$ Near the Superconductor-Insulator
Transition" }
\author{T. Schneider}
\address{Physik-Institut der Universit\"{a}t Z\"{u}rich, Winterthurerstrasse 190,\\
CH-8057 Z\"{u}rich, Switzerland}
\maketitle

\bigskip

Zuev \emph{et al}.\cite{zuev} reported measurements of the in-plane
magnetic
penetration depth of supposedly underdoped films of YBa$_{2}$Cu$_{3}$O$%
_{6+x} $, with $T_{c}$ from $6$ to $50$ K. They argue that their
data is inconsistent with a linear relationship between $T_{c}$ and
$1/\lambda _{ab}\left( T=0\right) $ as expected for a two
dimensional quantum superconductor to insulator (2D-QSI) transition
where $T_{c}$ vanishes and
the anisotropy $\gamma =\xi _{ab}/\xi _{c}$ diverges \cite%
{polen,book,tsphysb,parks}. $\xi _{ab,c}$ denote the correlation
length in the $ab$- plane and along the $c$- axis. In the following
we show that this interpretation is misleading because in these
films the 2D - limit was not attained.

\begin{figure}[tbp]
\centering
\includegraphics[totalheight=8cm]{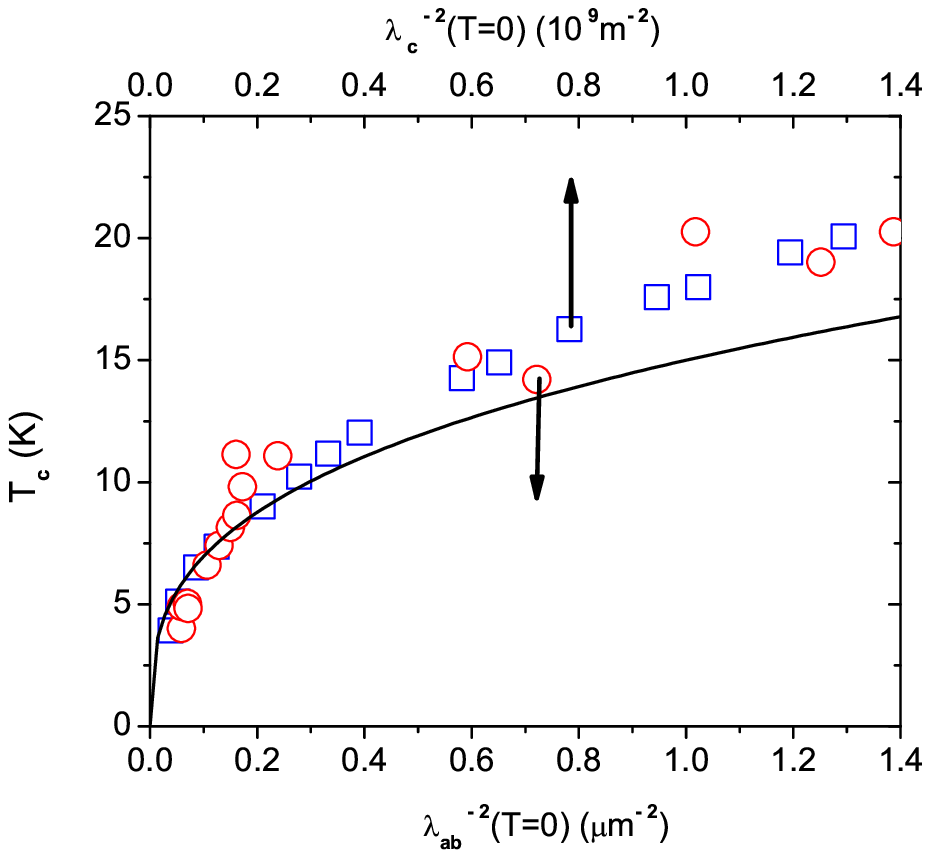}
\caption{$T_{c}$ \textit{vs.} $1/\lambda _{ab}^{2}\left( T=0\right)
$ for
underdoped YBa$_{2}$Cu$_{3}$O$_{7-\delta }$ films taken from Zuev \emph{et al%
}.\protect \cite{zuev} and $T_{c}$ \textit{vs.} $1/\lambda
_{c}^{2}\left( T=0\right) $ for underdoped
YBa$_{2}$Cu$_{3}$O$_{7-\delta }$ single crystals taken from Hosseini
\emph{et al}. \protect\cite{hosseini}. The solid line is
$T_{c}=15\left( 1/\lambda _{c}^{2}\left( T=0\right) \right)
^{1/3}$.} \label{fig1}
\end{figure}

It is well established that sufficiently homogeneous cuprates
undergo in the underdoped regime a 3D-2D crossover and close to the
underdoped limit a
2D-QSI transition where $T_{c}$ vanishes \cite%
{polen,book,tsphysb,parks,tshknjp,tsphsol}. At this quantum
transition the universal relation
\begin{equation}
T_{c}=\frac{\Phi _{0}^{2}d_{s}}{16\pi ^{3}k_{B}R}\frac{1}{\lambda
_{ab}^{2}\left( 0\right) },  \label{eq1}
\end{equation}%
holds, provided that the 2D limit is attained. $R\approx 2$ is a
universal number and $d_{s}$ the thickness of the independent
sheets. The 2D regime is entered when the anisotropy $\gamma =\xi
_{ab}/\xi _{c}$ becomes sufficiently large. Close to the 2D-QSI
transition $\gamma $ diverges as
\cite{book,tsphysb,parks,tshknjp,tsphsol}
\begin{equation}
\gamma \left( x\right) =\frac{\xi _{ab}\left( x\right) }{\xi
_{c}\left( x\right) }=\frac{\overline{\xi _{ab}}\left(
x-x_{u}\right) ^{-\overline{\nu }}}{d_{s}-b\left( x-x_{u}\right) },
\label{eq2}
\end{equation}%
because the in-plane correlation length diverges as $\xi
_{ab}=\overline{\xi _{ab}}\left( x-x_{u}\right) ^{-\overline{\nu
}}$, while the c- axis correlation length\ $\xi _{c}\left( x\right)
=d_{s}-b\left( x-x_{u}\right) $ tends to $d_{s}$. $x$ is the hole
concentration and the 2D-QSI transition occurs at $x_{u}\approx
0.05$. This divergency also implies that the c-axis penetration
depth diverges at this transition as $1/\lambda _{c}^{2}\left(
0\right) $ $\propto T_{c}^{(z+2/z}\propto T_{c}^{3}$ \cite{tshknjp}.
In the range $T_{c}\lesssim 10$ K this prediction is well confirmed
by the c-axis penetration depth data of heavily underdoped
YBa$_{2}$Cu$_{3}$O$_{7-\delta }$ single crystals of Hosseini
\emph{et al. } \cite{tshknjp,hosseini}. Thus, an essential
ingredient of the 2D-QSI transition is the divergency of the
anisotropy $\gamma =\xi _{ab}/\xi _{c}=\lambda _{c}/\lambda _{ab}$.
In Fig.\ref{fig1} we show $T_{c}$ \textit{vs.} $1/\lambda
_{ab}^{2}\left( T=0\right) $ for the YBa$_{2}$Cu$_{3}$O$_{7-\delta
}$ films of Zuev \emph{et al}. \cite{zuev} and $T_{c}$ \textit{vs.}
$1/\lambda _{c}^{2}\left( T=0\right) $ for underdoped
YBa$_{2}$Cu$_{3}$O$_{7-\delta }$ single crystals taken from Hosseini
\emph{et al}. \cite{hosseini}. The near coincidence of the data
implies that in these films $\gamma $ is essentially independent of
$T_{c}$ and adopts the value $\gamma \simeq 32$, which is very far
from the 2D-limit. Accordingly, the data of Zuev \emph{et
al}.\cite{zuev} do not fall onto the regime where the linear
relationship between $T_{c}$ and $1/\lambda _{ab}\left( T=0\right) $
(Eq.(\ref{eq1})) applies, because the observed reduction of $T_{c}$
is not associated with an increase of the anisotropy.

\end{document}